\documentclass[twocolumn,nofootinbib,amsmath,amssymb,aps,prd]{revtex4-1}

\usepackage{graphicx}
\usepackage{subcaption}

\usepackage{amsmath,amssymb}
\usepackage{amsfonts,amssymb,mathrsfs}

\usepackage[bookmarks=false,pdfstartview=FitH]{hyperref}
\usepackage[all]{hypcap}

\def\be{\begin{equation}}
\def\ee{\end{equation}}
\def\nn{\nonumber}
\def\f{\frac}
\def\tf{\tfrac}
\def\pl{{\rm Pl}}
\def\lp{\ell_\pl}

\def\d{\dot}

\def\t{\tilde}

\def\bra{\langle}
\def\ket{\rangle}
\def\dd{{\rm d}}

\def\de{\delta}

\def\ga{\gamma}

\def\mO{\mathcal{O}}

\def\mC{\mathcal{C}}

\def\mF{\mathcal{F}}
\def\lo{\ell_o}

\usepackage{color}

\begin{document}

\pagestyle{plain}

\title{A quantum gravity extension to the Mixmaster dynamics}

\author{Edward Wilson-Ewing} \email{edward.wilson-ewing@unb.ca}
\affiliation{Department of Mathematics and Statistics, University of New Brunswick, 
Fredericton, NB, Canada E3B 5A3}

\begin{abstract}

In the loop quantum cosmology effective dynamics for the vacuum Bianchi type I and type IX space-times, a non-singular bounce replaces the classical singularity.  The bounce can be approximated as an instantaneous transition between two classical vacuum Bianchi I solutions, with simple transition rules relating the solutions before and after the bounce.  These transition rules are especially simple when expressed in terms of the Misner variables: the evolution of the mean logarithmic scale factor $\Omega$ is reversed, while the shape parameters $\beta_\pm$ are unaffected.  As a result, the loop quantum cosmology effective dynamics for the vacuum Bianchi IX space-time can be approximated by a sequence of classical vacuum Bianchi I solutions, following the usual Mixmaster transition maps in the classical regime, and undergoing a bounce with this new transition rule in the Planck regime.

\end{abstract}

\maketitle

\section{Introduction}
\label{s.intro}

According to the Belinski-Khalatnikov-Lifshitz (BKL) conjecture \cite{Belinski:1969, Belinski:1970ew}, as a space-like singularity is approached in general relativity, at a generic point spatial derivatives become negligible compared to time-like derivatives and the equations of motion at neighbouring points decouple.  In this limit, the dynamics at each point become those of a Bianchi space-time, typically Bianchi VIII or IX, the Bianchi space-times with the richest dynamics.

Of course, close to a singularity quantum gravity effects are expected to become important and general relativity can no longer be trusted.  But if the BKL conjecture is correct, it suggests that understanding the nature of quantum gravity effects in the Bianchi space-times may provide important insights also for generic regions of space-time close to surfaces where general relativity would predict the formation of space-like singularities.  As such, a natural first step in studying quantum gravity effects is to start with the Bianchi models, and particularly the Bianchi type VIII and type IX space-times.

Loop quantum cosmology (LQC) is one approach to studying quantum gravity effects in cosmological space-times, based on a non-perturbative quantization of symmetry-reduced cosmological space-times following, as closely as possible, loop quantum gravity; for reviews see, e.g., \cite{Ashtekar:2011ni, Banerjee:2011qu}.  In particular, Bianchi type I, II and IX space-times have all been studied in LQC \cite{Bojowald:2003md, Chiou:2007sp, MartinBenito:2008wx, Ashtekar:2009vc, Ashtekar:2009um, WilsonEwing:2010rh, Singh:2013ava}.

In LQC, semi-classical states (i.e., states that at late times are sharply peaked around a classical solution) are not only interesting from a physical point of view, but also have simple dynamics in the sense that quantum fluctuations do not grow significantly so long as the spatial volume of the space-time always remains much larger than the Planck volume (or in the anisotropic case, that all scale factors remain much larger than the Planck length $\lp$) \cite{Rovelli:2013zaa}.  Since quantum fluctuations are small, it is reasonable to approximate $\bra \mO^2 \ket \approx \bra \mO \ket^2$ for any observable $\mO$.  The dynamics of the expectation values of observables are given by some `effective equations', that in the classical limit are identical to the equations of motion of general relativity, but also include quantum corrections that become important when the space-time curvature nears the Planck scale \cite{Ashtekar:2006wn, Taveras:2008ke}.  In the cases where numerical solutions of the quantum theory have been derived, the effective equations closely track the dynamics of sharply peaked quantum states, see, e.g., \cite{Ashtekar:2006wn}.

While the effective LQC dynamics have not yet been compared with full quantum dynamics for the Bianchi space-times (although see \cite{MartinBenito:2009qu, Diener:2017lde, Pawlowski-talk} for numerical studies working, in part, towards this), so long as the three scale factors remain much larger than the Planck length, the observables in the quantum theory (i.e., the scale factors and their conjugate momenta) will be heavy degrees of freedom; in this case quantum fluctuations can safely be neglected for semi-classical states and the effective dynamics are expected to track the quantum dynamics of semi-classical states~\cite{Rovelli:2013zaa}.  Numerical studies of the effective equations for Bianchi space-times have found that quantum gravity effects generate a non-singular bounce that replaces the big-bang singularity \cite{Gupt:2012vi, Corichi:2012hy, Corichi:2015ala}, and quantum gravity effects are important only for a few $t_{\rm Pl}$.  Even a short time away from the bounce, the solution is extremely well approximated by classical general relativity, so the LQC bounce can be treated as an instantaneous transition between two classical solutions.  As shall be shown, there are simple transformation rules describing how, in the effective LQC dynamics, the classical Bianchi solutions on either side of the bounce are related.

These transformation rules for the LQC bounce are analogous to the transition rules of the Mixmaster dynamics that describe the evolution of the vacuum Bianchi IX space-time in general relativity, and in fact provide a quantum gravity extension to them: the Mixmaster transition rules describe the classical evolution, while the LQC transition rules describe the dynamics, with quantum gravity corrections, of the Bianchi IX space-time near the Planck scale.

The vacuum Bianchi I and Bianchi IX solutions in general relativity are briefly reviewed in Sec.~\ref{s.gr}.  The LQC bounce transition rules are derived for the vacuum Bianchi I space-time in Sec.~\ref{s.lqc}, and these results are extended to the vacuum Bianchi IX space-time in Sec.~\ref{s.mixmaster}.

\newpage

\section{Classical Solutions}
\label{s.gr}

The line element for the Bianchi I space-time is
\be
\dd s^2 = - N^2 \dd t^2 + e^{2 \alpha_1} \dd x_1^2 + e^{2 \alpha_2} \dd x_2^2 + e^{2 \alpha_3} \dd x_3^2,
\ee
where the $\alpha_i$ are the logarithms of directional scale factors, $a_i = e^{\alpha_i}$, and are functions of $t$ only.  The lapse $N$ is also a function of $t$ only.

The dynamics for the space-time can be determined from the Hamiltonian constraint of the Arnowitt-Deser-Misner formulation of general relativity \cite{Arnowitt:1962hi}.   These dynamics have a particularly simple form when the lapse is chosen to be $N = V = \exp(\sum_i \alpha_i)$, in this case the Hamiltonian constraint for the vacuum Bianchi I space-time in general relativity is
\be \label{ham-cl}
\mC_I = \f{\lo^{-3}}{32 \pi G} \Big[ \Pi_1^2 + \Pi_2^2 + \Pi_3^2 - 2 (\Pi_1 \Pi_2 + \Pi_1 \Pi_3 + \Pi_2 \Pi_3) \Big].
\ee
Here the $\Pi_i(t)$ are canonically conjugate to $\alpha_i$,
\be
\{\alpha_i, \Pi_j\}  = -8 \pi G \, \de_{ij},
\ee
and classically, e.g., $\Pi_1 = a_1 a_2 \d a_3 + a_1 \d a_2 a_3$ with the dot denoting a derivative with respect to proper time (i.e., $\d f = N^{-1} df/dt$).  Here $\lo^3$ is the spatial volume with respect to the coordinates $x_i$; for non-compact space-times it is necessary to restrict integrals over the homogeneous spatial slice to a fiducial cell and then $\lo^3$ is the volume of the fiducial cell.  For Bianchi IX, a convenient coordinate choice gives $\lo^3 = 16 \pi^2$.

The dynamics are given by $\dd\mO/\dd\tau = \{\mO, \mC\}$, and for the Hamiltonian \eqref{ham-cl} all $\Pi_i$ are constants of the motion while
\be \label{cl-alpha}
\alpha_i = \f{\Pi_j + \Pi_k - \Pi_i}{2 \lo^3} \, \tau + \alpha_i^{(0)},
\ee
with $i, j, k$ all different, $\alpha_i^{(0)}$ a constant of integration determined by the initial conditions, and $\tau$ the harmonic time coordinate for $N = V$.

It is convenient to express the logarithmic scale factors in terms of the mean logarithmic scale factor $\Omega = \sum_i \alpha_i / 3$ and the two shape parameters \cite{Misner:1969ae}
\be
\beta_+ = \f{1}{4} (\alpha_1 + \alpha_2 - 2 \alpha_3), \quad
\beta_- = \f{1}{2\sqrt{3}} (\alpha_1 - \alpha_2).
\ee
The dependence of $\Omega$ and $\beta_\pm$ on $\tau$ follows from \eqref{cl-alpha}.  Clearly, the trajectory of the Bianchi I solution in the 3-dimensional $(\Omega, \beta_\pm)$ space is a straight line.

\bigskip

For the Bianchi IX space-time, the metric (with $N=V$) is
\be
\dd s^2 = - V^2 \dd\tau^2 + \sum_i e^{2 \alpha_i} (\mathring{\omega}^i)^2,
\ee
with $\mathring{\omega}^i$ a 1-form satisfying $\dd \mathring{\omega}^i = \tf{1}{2} \epsilon^i{}_{jk} \mathring{\omega}^j \wedge \mathring{\omega}^k$.  The Hamiltonian constraint has an additional potential term compared to \eqref{ham-cl} due to the presence of spatial curvature: $\mC_{IX} = \mC_I + U(\alpha_i)$, with $\mC_I$ the Bianchi I Hamiltonian constraint \eqref{ham-cl} and the dominant terms in $U$ are
\be
U \sim \f{\lo^3}{32 \pi G} e^{4 \Omega} \Big( e^{4 \beta_+ + 4 \sqrt{3} \beta_-} + e^{4 \beta_+ - 4 \sqrt{3} \beta_-} + e^{-8 \beta_+} \Big).
\ee
There are also additional terms in the potential, but these terms, near the singularity, are negligible for generic Bianchi IX solutions \cite{Ringstrom:2000mk} (the same terms dominate the potential for generic vacuum Bianchi VIII solutions near the singularity \cite{Brehm:2016cck}.)  This potential has a triangular symmetry in the $\beta_\pm$ plane, and as the three walls of the potential are exponentially steep, the potential walls can be approximated by a hard wall (located where $U \sim 1/G$) that the system bounces off instantaneously in the $(\Omega, \beta_\pm)$ space.  In this approximation, away from the potential walls the system follows a Bianchi I solution, and when the system bounces off one of the potential walls, it instantaneously transitions from one Bianchi I solution to another, with the new solution determined from the previous one by simple transition rules: two $\Pi_i$ remain unchanged while the third (depending on which of the three exponential walls in $U$ the system bounces off) transforms as
\be \label{cl-trans}
\Pi_j \to \tilde \Pi_j = 2 \Pi_k + 2 \Pi_l - \Pi_j,
\ee
with $j,k,l$ all different.  $\Omega, \beta_\pm$ are continuous in $\tau$, though not differentiable at the transition times.  For details see, e.g., \cite{Belinski:1969, Belinski:1970ew, Misner:1969ae, Montani:2007vu, Uggla:2013laa, Berger:2014tev, Wilson-Ewing:2017vju}.

As the singularity is approached, $\Omega \to -\infty$ monotonically while $\beta_\pm$ repeatedly bounce off the walls of the triangular potential $U$.  This triangle becomes larger as $\Omega$ decreases due to the $e^{4\Omega}$ prefactor in $U$, so the Mixmaster dynamics (i.e., the dynamics of the vacuum Bianchi IX space-time as it approaches the $V=0$ singularity) can be seen as a particle in the $\beta_\pm$ plane bouncing off the walls of an expanding triangular potential.

Alternately, the Mixmaster dynamics can also be viewed as a particle in the three-dimensional $(\Omega, \beta_\pm$) space with the potential walls forming a bottomless triangular pyramid.  In the approach to the singularity, the system continually moves towards the singularity at $\Omega \to -\infty$, bouncing off the pyramidal walls an infinite number of times before reaching the singularity.

\section{The LQC Bounce: Bianchi I}
\label{s.lqc}

Numerical solutions of the LQC effective dynamics of the Bianchi I space-time show that a non-singular bounce occurs very rapidly, and that quantum gravity effects quickly become negligible either side of the bounce---to an excellent approximation either side of the bounce can be described by a classical Bianchi I solution \cite{Gupt:2012vi}.  Therefore, the LQC bounce of the Bianchi I space-time can be approximated as an instantaneous transition between two classical Bianchi I solutions, much as how the Mixmaster dynamics can be approximated by a sequence of Bianchi I solutions linked by instantaneous transitions.  

The Hamiltonian constraint
\begin{align} \label{lqc-b1}
\mC_I^{(LQC)} = & \, -\f{V^2 \lo^{-3}}{8 \pi G \ga^2 \Delta}
\Big[ \sin \mF_1 \sin \mF_2 + \sin \mF_1 \sin \mF_3 \nn \\ & \qquad \qquad \qquad \qquad
+ \sin \mF_2 \sin \mF_3 \Big]
\end{align}
generates the effective LQC dynamics for the vacuum Bianchi I space-time \cite{Chiou:2007sp, Ashtekar:2009vc}.  Here
\be
\mF_i = \f{\ga \sqrt\Delta}{2 V} (\Pi_j + \Pi_k - \Pi_i),
\ee
with $i,j,k$ all different, while $\Delta \sim \lp^2$ is the smallest non-zero eigenvalue of the area operator of loop quantum gravity, and $\gamma$ is the Barbero-Immirzi parameter.  For details on the quantum theory, see \cite{Ashtekar:2009vc}.

One way to determine how the two classical solutions either side of the LQC bounce are related is to notice that the equations of motion for all of the $\Pi_i$ in the effective LQC dynamics are identical: $\dd \Pi_1 / \dd\tau = \dd \Pi_2 / \dd\tau = \dd \Pi_3 / \dd\tau$.  This is because $\mC_I^{(LQC)}$ depends on the $\alpha_i$ only through $V = \exp (\sum_i \alpha_i)$, and the Poisson bracket $\{\Pi_i, V\} = 8 \pi G V$ is the same for all $\Pi_i$, so
\be \label{dpi}
\f{\dd \Pi_i}{\dd\tau} = \{\Pi_i, \mC_I^{(LQC)}\} = 8 \pi G V \, \f{\de \mC_I^{(LQC)}}{\de V}.
\ee
Since the $\Pi_i$ are constant in the classical regime on either side away from the bounce, this implies the key result that during the bounce all of the $\Pi_i$ will be shifted by exactly the same amount: $\Pi_i \to \t\Pi_i = \Pi_i + \Delta\Pi$, with $\Delta\Pi$ given by the integral of \eqref{dpi} with respect to $\tau$ over the short period of time near the bounce that \eqref{dpi} is non-zero.  A simple way to calculate the value of $\Delta\Pi$ is by noting that the three $\Pi_i$ before the bounce must satisfy the classical Hamiltonian constraint \eqref{ham-cl}, and so must the $\t\Pi_i = \Pi_i + \Delta\Pi$ after the bounce, away from the small bounce region where quantum gravity effects are important and cannot be neglected.  Given the requirements that the $\Pi_i$ and $\Pi_i + \Delta\Pi$ both satisfy the classical constraint $C_I = 0$, the only possible solutions for $\Delta\Pi$ are $\Delta\Pi=0$ (the pre-bounce solution) and
\be
\Delta\Pi = - \f{2}{3} (\Pi_1 + \Pi_2 + \Pi_3),
\ee
for the post-bounce solution \cite{Wilson-Ewing:2017vju}.  It then follows that the values of the $\Pi_i$ either side of the LQC bounce, in the regions well-approximated by a classical solution, transform as
\be \label{pi}
\Pi_i \to \t\Pi_i = \Pi_i - \f{2}{3} (\Pi_1 + \Pi_2 + \Pi_3).
\ee
Note that $\sum_i \Pi_i \to \sum \t\Pi_i = -\sum_i \Pi_i$; this is a signature of the LQC bounce in the volume $V(\tau)$.  The transformation rule \eqref{pi} does not depend on the bounce occurring rapidly, but since the LQC bounce is nearly instantaneous \cite{Gupt:2012vi}, it is in addition possible to approximate the exact solution for $\alpha_i(\tau)$ by a piecewise constant function.

From the transition rule \eqref{pi}, it is possible to calculate how the logarithmic scale factors of the classical Bianchi I solutions approximating the LQC solution on either side of the bounce are related.  Given the classical solution \eqref{cl-alpha} on one side of the bounce, on the other side the solution is
\be
\t\alpha_i(\tau) = \f{\t\Pi_j + \t\Pi_k - \t\Pi_i}{2 \lo^3} \, \tau + \t\alpha_i^{(0)},
\ee
and from the transformation rule \eqref{pi}, it follows that
\be \label{alpha}
\t \alpha_i(\tau) = \alpha_i(\tau) - \f{\Pi_1 + \Pi_2 + \Pi_3}{3 \lo^3} (\tau - \tau_b),
\ee
where $\t\alpha_i^{(0)}$ has been chosen to ensure that $\alpha_i(\tau)$ is continuous at the bounce time $\tau_b$.

From the transformation rule \eqref{alpha}, it follows that
\begin{gather} \label{lqc-tr}
\t\Omega(\tau) = \Omega(\tau) - \f{\Pi_1 + \Pi_2 + \Pi_3}{3 \lo^3} \, (\tau - \tau_b), \\ \label{lqc-tr2}
\t \beta_\pm(\tau) = \beta_\pm(\tau),
\end{gather}
and their velocities also change in a simple manner:
\be \label{vel}
\f{\dd \t \Omega}{\dd\tau} = - \f{\dd \Omega}{\dd \tau}, \qquad
\f{\dd \t \beta_\pm}{\dd\tau} = \f{\dd \beta_\pm}{\dd \tau}.
\ee
The LQC bounce exactly reverses the evolution of the mean logarithmic scale factor $\Omega$, which changes from contraction to expansion with the amplitude $|\dd\Omega/\dd\tau|$ unchanged; on the other hand the dynamics of the shape parameters $\beta_\pm$ are entirely unaffected by the LQC bounce and continue evolving as before.  Note that the momenta conjugate to $\Omega, \beta_\pm$ given by $p_\Omega = -12\pi G \, \dd\Omega/\dd\tau$ and $p_\pm = 12\pi G \, \dd\beta_\pm/\dd\tau$ (see, e.g., \cite{Montani:2007vu}) therefore transform as $p_\Omega \to \t p_\Omega = -p_\Omega$ and $p_\pm \to \t p_\pm = p_\pm$.

Note that due to the symmetry of the Bianchi I space-time, where all directions are treated equally in both $\mC_I$ and $\mC_I^{(LQC)}$, the shape parameters must either (i) be unaffected by the bounce, or (ii) reverse direction after the bounce.  Anything else would require the presence of a preferred direction \cite{Uggla}.  It is the first possibility that occurs in LQC.  Note that this simple argument implies that either possibility (i) or (ii) is realized also in any other theory that gives a bounce in a Bianchi I space-time (with a good classical limit either side of the bounce) without introducing a preferred direction.

Finally, note that in the effective LQC dynamics for the Bianchi I space-time (the absolute value of) the expansion
\be \label{theta}
\theta = \f{1}{NV} \f{\dd V}{\dd\tau} = 3 e^{-3\Omega} \f{\dd \Omega}{\dd\tau}
\ee
is bounded above by the Planck scale \cite{Corichi:2009pp}, and the LQC bounce occurs when the expansion nears $\sim \lp^{-1}$ (the exact value of $\theta$ at the LQC bounce may depend on the solution), so the `potential wall' responsible for the LQC bounce is located at $\theta \sim \lp^{-1}$.  Importantly, the bounce can easily happen when all scale factors satisfy $a_i \gg \lp$ and the effective dynamics remain valid.  Since the expansion depends on $\dd\Omega/\dd\tau$ in addition to $\Omega$, for different values of $\dd\Omega/\dd\tau$ the LQC bounce will occur at different $\Omega$ and therefore the `potential wall' of the LQC bounce cannot be located at a universal value of $\Omega$ in the $(\Omega, \beta_\pm)$ space for all solutions.  This is different to the Mixmaster dynamics of general relativity, where the potential walls form the same bottomless triangular pyramid in the $(\Omega,\beta_\pm)$ space no matter the Bianchi IX solution.  Nonetheless, although the location of the LQC `potential wall' depends on both $\Omega$ and $\dd\Omega/\dd\tau$ (but not $\beta_\pm$), it always provides a `bottom wall' the LQC solution bounces off with simple transition rules relating the classical Bianchi I solutions before and after the bounce.

\section{Quantum Mixmaster Dynamics}
\label{s.mixmaster}

The LQC effective dynamics for the Bianchi IX space-time are generated by the Hamiltonian constraint \cite{Singh:2013ava}
\be
\mC_{IX}^{(LQC)} = \mC_I^{(LQC)} + U(\alpha_i),
\ee
with $\mC_I^{(LQC)}$ the LQC Hamiltonian constraint for the Bianchi I space-time \eqref{lqc-b1}, and the potential $U(\alpha_i)$ unchanged from the classical theory.  (There are some ambiguities in the quantization of Bianchi space-times with non-vanishing spatial curvature in LQC, see \cite{Singh:2013ava} for details.  This effective Hamiltonian corresponds to the `K' loop quantization and neglects inverse triad effects.)

\begin{figure*}
\begin{subfigure}[t]{0.55\textwidth} \vskip -20pt
\includegraphics[width=\textwidth]{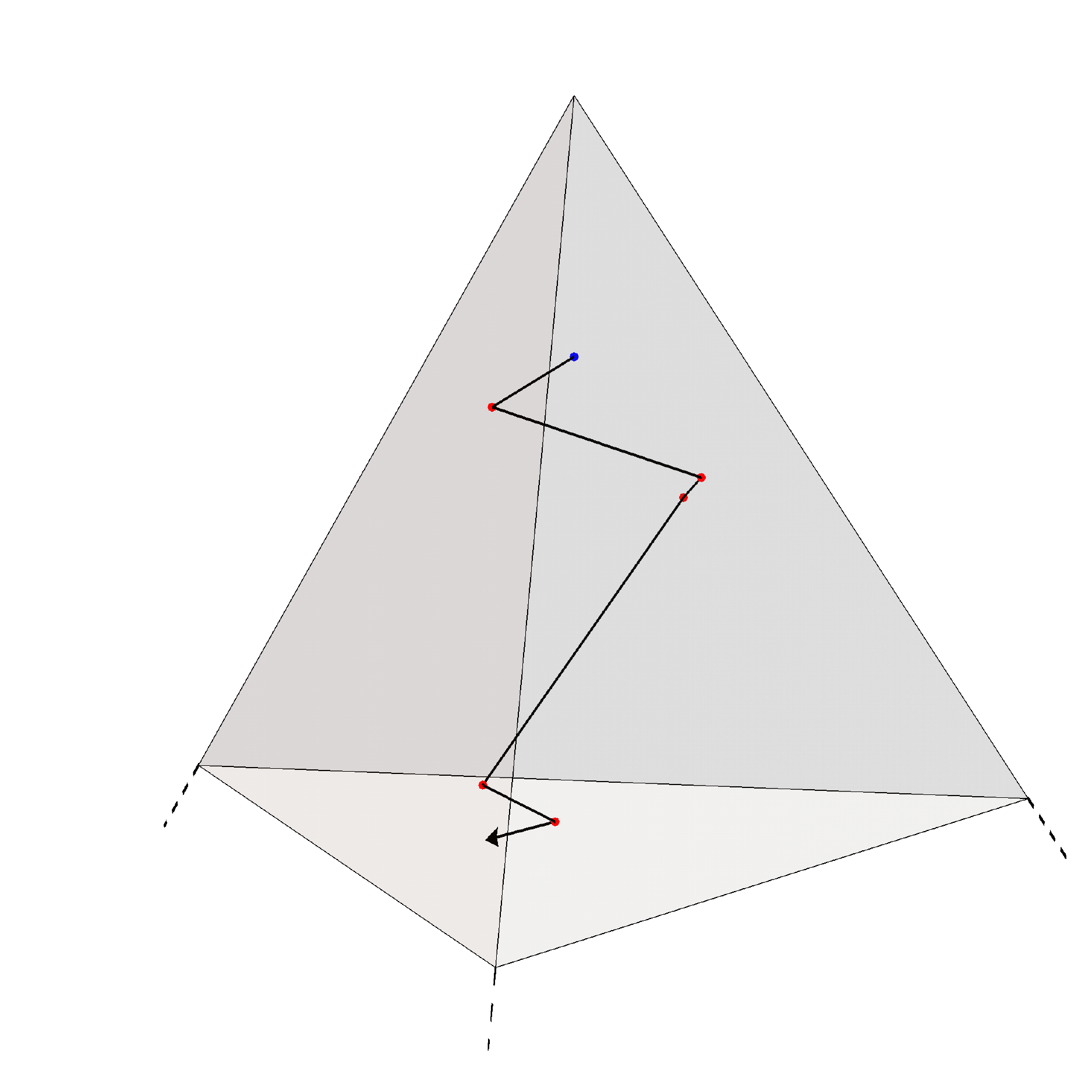}
\end{subfigure}%
\begin{subfigure}[t]{0.33\textwidth} \vskip -5pt
\includegraphics[width=\textwidth]{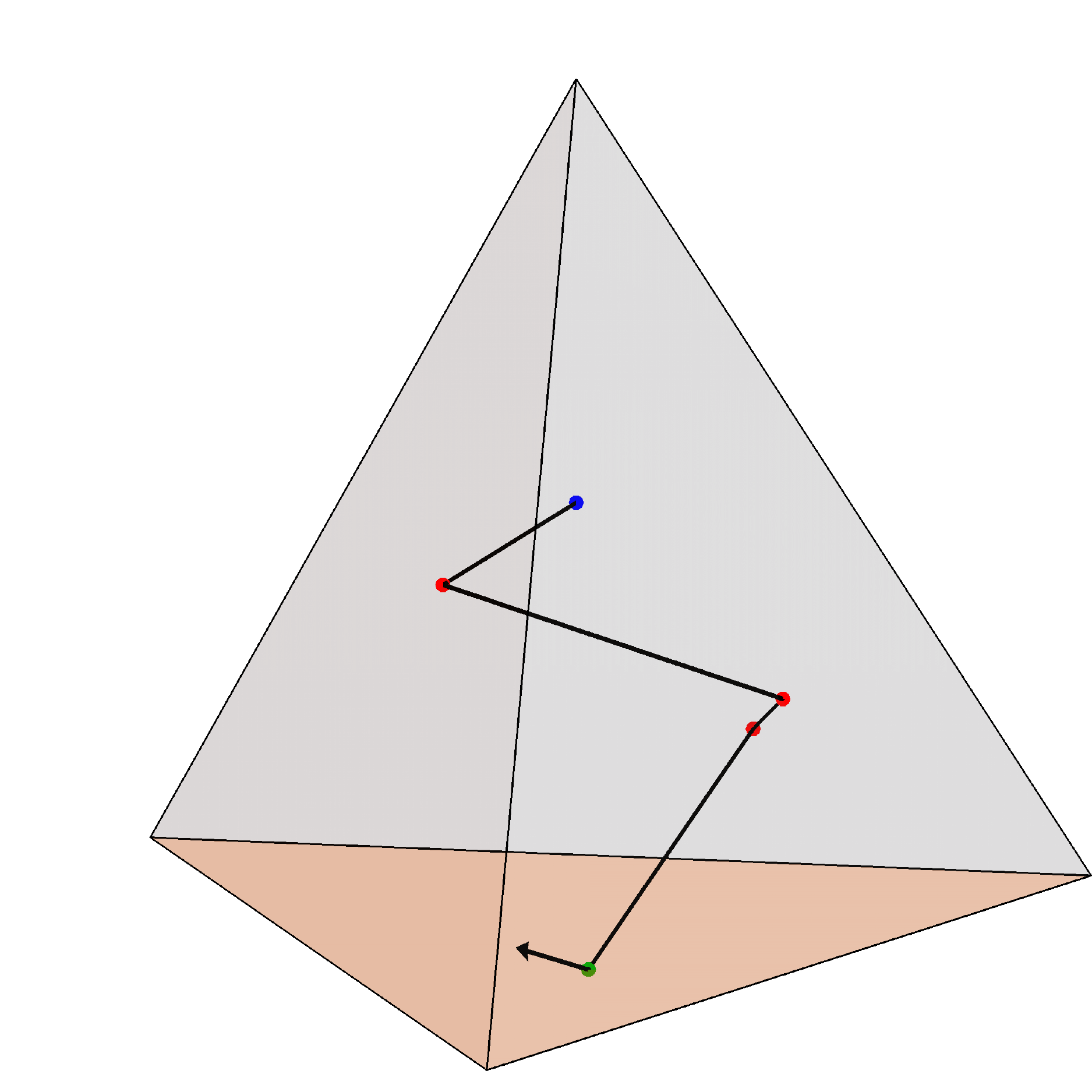}
\end{subfigure}
\caption{ \small
A schematic depiction of the classical Mixmaster dynamics is given on the left, and of the effective LQC Mixmaster dynamics on the right.  The vertical axis is $\Omega$ (with $\Omega$ increasing vertically) and the plane is spanned by $\beta_\pm$.  In this example, the initial conditions are given at the blue dot, and the transitions due to one of the three $U(\beta_\pm)$ potential walls are indicated by red circles.  In the classical Mixmaster dynamics, the triangular pyramid is bottomless so there are an infinite number of transitions as the singularity is approached.  On the other hand, in the effective LQC Mixmaster dynamics there is now a bottom floor due to quantum gravity effects; the LQC bounce off this floor is indicated in the figure on the right by a green circle.  Following the LQC bounce, with the volume $V$ now increasing, the system will continue to undergo the usual classical Kasner transitions whenever the system hits one of the spatial curvature walls.}
\label{fig}
\end{figure*}

Away from the LQC bounce, the classical dynamics are an excellent approximation to the LQC effective dynamics, and these dynamics, as reviewed in Sec.~\ref{s.gr}, are in turn well approximated by a sequence of Bianchi I solutions where the spatial curvature (and therefore the potential $U$) is negligible except during the transitions between Bianchi I solutions.

These transitions are very rapid and can be approximated as being instantaneous.  Since the LQC bounce also occurs very rapidly in the effective theory (numerical simulations find $\sim t_{\rm Pl}$ for the Bianchi I space-time \cite{Gupt:2012vi}), it is also reasonable to approximate the LQC bounce as being instantaneous.  In this case, absent fine-tuned initial conditions, it is reasonable to expect that: (i) LQC effects will be negligible during bounces off the potential $U$, and (ii) the potential $U$ will be negligible during the LQC bounce.  Then, the rules \eqref{cl-trans} derived classically for the Mixmaster transitions off the potential walls of $U$, and those derived for the Bianchi I LQC bounce \eqref{pi}, would both remain the same for the effective LQC Mixmaster dynamics.  It would be good to check the assumptions (i) and (ii) above by numerically solving, for a wide range of initial conditions, the effective LQC dynamics for the Bianchi IX space-time; this is left for future work.

Based on these two assumptions, the LQC Mixmaster dynamics can be described as a sequence of classical Bianchi I solutions bouncing off the potential walls with $\Omega$ decreasing until the LQC bounce occurs.  At the LQC bounce, $\Omega$ is reflected following \eqref{lqc-tr}, while the shape parameters are entirely unaffected by the LQC bounce.  The LQC bounce is expected to occur when (the absolute value of) the expansion reaches the Planck scale, $|\theta| \sim \lp^{-1}$; as can be seen in \eqref{theta}, this occurs at
\be
e^{3\Omega} \sim 3 \, \lp \, \left| \f{\dd \Omega}{\dd\tau} \right|.
\ee
After the bounce, now with $\Omega$ increasing, the dynamics is approximated by another sequence of classical Bianchi I solutions, again bouncing off the potential walls of $U$ following the classical transition rules reviewed in Sec.~\ref{s.gr}.  This picture holds if the spatial curvature is negligible during the LQC bounce; if $U$ cannot be neglected during the LQC bounce, more work is necessary to determine how $\Omega$ and $\beta_\pm$ transform in this case.

These LQC Mixmaster dynamics are piecewise linear in the $(\Omega,\beta_\pm)$ space, with bounces off a triangular pyramid with the three usual Mixmaster spatial curvature `upper walls' and a new quantum gravity `bottom wall', this is depicted in Fig.~\ref{fig}.

Alternately, the LQC Mixmaster dynamics can be projected on the $\beta_\pm$ plane, where the trajectory is again piecewise linear with the system bouncing off the triangular potential walls of $U$.  In a contracting space-time the potential walls will initially be moving away from the origin, but when the LQC bounce occurs the potential walls will reverse direction and move back towards the origin.  During the LQC bounce, the trajectory of the system in the $\beta_\pm$ plane is unchanged as seen in \eqref{lqc-tr2}.

Finally, the transition rules for the LQC bounce in the Mixmaster model can also be expressed in terms of other variables, for example the Kasner exponents transform as $k_i \to \t k_i = 2/3 - k_i$, and the BKL $u$ parameter transforms as $u \to \t u = (u+2)/(u-1)$, see \cite{Wilson-Ewing:2017vju} for details.

\section{Discussion}
\label{s.disc}

For the vacuum Bianchi IX space-time, the LQC effective dynamics provides a quantum gravity extension to the Mixmaster dynamics by introducing a new type of transition that occurs when the expansion $\theta$ reaches the Planck scale.  This effectively introduces a new `quantum gravity bottom' to the pyramid-shaped potential in the $(\Omega,\beta_\pm)$ space; the location of this LQC potential wall depends on $\Omega$ and $\dd\Omega/\dd\tau$, but not $\beta_\pm$.  Alternately, the LQC bounce can be viewed in the $\beta_\pm$ plane as the reversal of the motion of the potential walls (from initially moving away from the origin to afterwards moving back towards the origin at the same speed) without having any impact on the dynamics of the shape parameters $\beta_\pm$.

The classical Mixmaster dynamics are known to be chaotic \cite{Barrow:1981sx, Chernoff:1983zz, Cornish:1996yg}, and the LQC Mixmaster dynamics are essentially the same, except with a new additional type of transition.  It seems likely the LQC Bianchi IX dynamics will also be chaotic for the reason that there will be an infinite number of expansion-recollapse-contraction-bounce cycles, and each expansion-recollapse-contraction segment is identical to a portion of the classical Bianchi IX space-time dynamics.  Since the LQC bounce does not reverse (or change in any way) the dynamics of $\beta_\pm$, it seems the sensitivity to initial conditions and the mixing of solutions generated during the expansion-recollapse-contraction segment will remain present and add up over consecutive cycles, in which case the dynamics can be expected to be chaotic.  For more on this point, see \cite{Wilson-Ewing:2017vju}.

In earlier work, on the other hand, it was suggested that inverse volume effects could remove the chaotic behaviour from the Bianchi IX space-time \cite{Bojowald:2003xe}.  However, this was based on the assumption that the Bianchi IX space-time would contract indefinitely; this assumption is violated by the occurrence of the bounce in LQC.  In future work, it would be interesting to extend the results obtained here to include inverse triad effects (that were assumed to be negligible here); these are expected to become important only for a Bianchi IX space-time that has at least one scale factor reach (at some time) $\sim \lp$.

Note that the above argument indicating that the effective LQC dynamics for the Bianchi IX space-time may be chaotic is not relevant for the BKL conjecture: the BKL conjecture, if correct, will presumably only hold for a short time when the curvature is large, while the argument for chaos in the Bianchi IX space-time relies in part on an infinite sequence of transitions between Bianchi I solutions.  If, as expected, the BKL behaviour only lasts a short time before the system bounces and leaves the BKL regime, the finite number of transitions between Bianchi I solutions may not be sufficient to cause chaos.

It is interesting to add a massless scalar field to the Bianchi space-times.  This gives an extra contribution to the Hamiltonian constraint, and some other relations are modified (see \cite{Wilson-Ewing:2017vju} for details), but the transition rules \eqref{lqc-tr}--\eqref{vel} remain the same: they are not affected by the presence of the massless scalar field.

Finally, another possibility would be to add a cosmological constant and/or other matter fields like radiation or dust; these would be expected to affect the dynamics especially in the classical regime, away from the bounce that was the focus here.  See \cite{Barrow:2017yqt} for work along these lines in a different realization of a cyclic Mixmaster space-time where the non-singular bounce is caused by a ghost field rather than quantum gravity effects; similar qualitative results (insofar as the classical regime is concerned) would likely hold in LQC.

\newpage

\noindent
{\it Acknowledgments:} 
I thank Claes Uggla for very helpful discussions and comments on an earlier draft of the paper, and Marco de Cesare for his help in preparing the figures.
This work was supported in part by the Natural Science and Engineering Research Council of Canada.

\small
\raggedright

\end{document}